\documentclass[final,5p,times,twocolumn]{elsarticle}

\usepackage{amssymb}
\usepackage{longtable}
\usepackage{bm}
\usepackage{multirow}
\usepackage{hyperref}
\usepackage{ulem}
\hypersetup{colorlinks,citecolor=blue,filecolor=blue,linkcolor=blue,urlcolor=blue}
\usepackage{color}
\usepackage{dcolumn}
\usepackage{tabularx}
\usepackage[usenames,dvipsnames]{xcolor}

\definecolor{awesome}{rgb}{1.0, 0.13, 0.32}
\newcommand{\iso}[2]{$^{#1}$#2}

\newcommand{\BoxScore}[0]{\textsf{BoxScore}}

\journal{}

\begin{document}

\begin{frontmatter}


\title{BoxScore - A real-time beam-diagnosis program for CAEN digitizer x730 series}


\author[inst1]{T.~L.~Tang}
\ead{rtang@fsu.edu}
%\author[inst1]{C.~R.~Hoffman}
%\ead{crhoffman@anl.gov}
\affiliation[inst1]{organization={Physics Division, Argonne National Laboratory},
            addressline={9700 S Cass Ave}, 
            city={Lemont},
            postcode={60439}, 
            state={Illinois},
            country={USA}}

\begin{abstract}

\BoxScore~is a real-time beam diagnosis and monitoring program for the CAEN x730 series digitizer that was developed for the ATLAS in-flight system at Argonne National Laboratory.
The CAEN x730 series digitizer, with built-in Digital Pulse Processing for the Pulse-Height-Analysis, can analyze the input signal in real-time using a trapezoidal filter. % a powerful device for many nuclear physics applications. 
%\crhnote{the sentence above needs `traces' and maybe even 'trapezoidal filter' defined. For example, should traces be input signals, and the trapezoidal filter does what?}
\BoxScore reads the digitizer's buffer directly, builds and saves events to local files, plots filled histograms for particle identification, and outputs the rates of selected isotopes every second. Implementation of \BoxScore has shortened the time needed for in-flight beam-tuning and has potential applications for other nuclear physics experiments.
\end{abstract}



\begin{keyword}
%keyword one \sep keyword two

%CAEN x730 Digitizer \sep Real-time beam-diagonsis
%\crhnote{I think that there are certain approved keywords only but maybe I am wrong}

\end{keyword}

\end{frontmatter}


\section{Introduction}

The CAEN x730 series digitizers are powerful data acquisition devices for many nuclear physics applications. 
In particular, the DT5730 8-channel digitizer \cite{DT5730} using USB connection (30 MB/s) and the V1730 16-channel digitizer \cite{V1730} using optical fiber connection (80 MB/s) are used at Argonne National Laboratory. 
The built-in Digital Pulse Processing for the Pulse Height Analysis (DPP-PHA) function provides a real-time trapezoid filter~\cite{Jordanov1994} for pulse height and pile-up detection. 
The digitizer gives the essential information (pulse height and timestamp) that can be used widely in nuclear physics applications such as gamma-ray spectroscopy, radiation imaging, and particle identification (PID).
%\crhnote{Is this true? What role does the leading edge discriminator play then in determining the timestamp etc? I thought leading edge disc. determined time, trap. filter does energy (also perhaps a reference to the math involved in the trap filter would be good, i know there is one somewhere)}

The Argonne Tandem Linear Accelerator Systems (ATLAS) User Facility has been involved in the production of radioactive ion beams via the in-flight technique for a number of years~\cite{Harss2000}. In 2019, the ATLAS in-flight system was upgraded through a number of hardware improvements including a dedicated magnetic separator, the Argonne in-flight radioactive ion-beam separator (RAISOR) \cite{RAISOR}. As a consequence of the increased complexity involved with the new installations, traditional beam optimization techniques were limited and time consuming. Hence, a need arose for a real-time beam diagnostic and monitoring system.
\mbox{\BoxScore}~\cite{BoxScore} is a program that was built to remedy this situation through the use of a single CAEN x730 series digitizer in conjunction with a series of silicon-based charged particle (sensor) detector systems.
The program is written in C++, and based upon the following readily available software libraries: CAENDigitizer 2.12+, CAENComm 1.4+, \mbox{CAENVMELib 2.5+}, CAEN A3818 Driver 1.6.4 (for PCI optical link), and \mbox{CERN ROOT 6+}~\cite{ROOT6}. 
The program has been developed from an example that comes with the CAENDigitizer library. 

The primary goal for \BoxScore~is beam diagnosis via charged particle identification (PID) and the corresponding isotope rate determinations using a $\Delta E-E$ telescope comprised of silicon detectors~\cite{Lilley2001}. 
A perpendicular beam profile (or x-y plane image) is also possible with a spatially sensitive detector. %\crhnote{This needs a good reference or need to define what is meant by diagnosis, i.e. rate determination, energies, etc...}
There are two modes of operation for the x730 digitizer under the DPP-PHA firmware: 1) the ``list" mode provides the pulse height and timestamp of the triggered signal.
2) the ``mixed" mode outputs the waveform, pulse height, and timestamp.  
\BoxScore~can be a multi-channel oscilloscope in the ``mixed" mode.
The program builds events from raw data based on a given time window, displays the events in histograms, and saves the data and histograms into a CERN ROOT file. 
Isotope selections can be done online and graphically using the \texttt{TCutG} Class by CERN ROOT for picking various isotopes in the PID plot. 
The rates (counts per second) of the selected isotopes can be outputted to a local or remote database. %\crhnote{trigger rates needs to be defined, or maybe say just the channel rates?}
An open-source database InfluxDB~\cite{influx} was used to store the rates for recording, sharing, and monitoring.

%----------------------FIGURE 1-----------------
\begin{figure}[ht]
\centering
%trim= left bottom right top
\includegraphics[trim=5cm 7.5cm 7cm 6cm, clip, width=8.5cm]{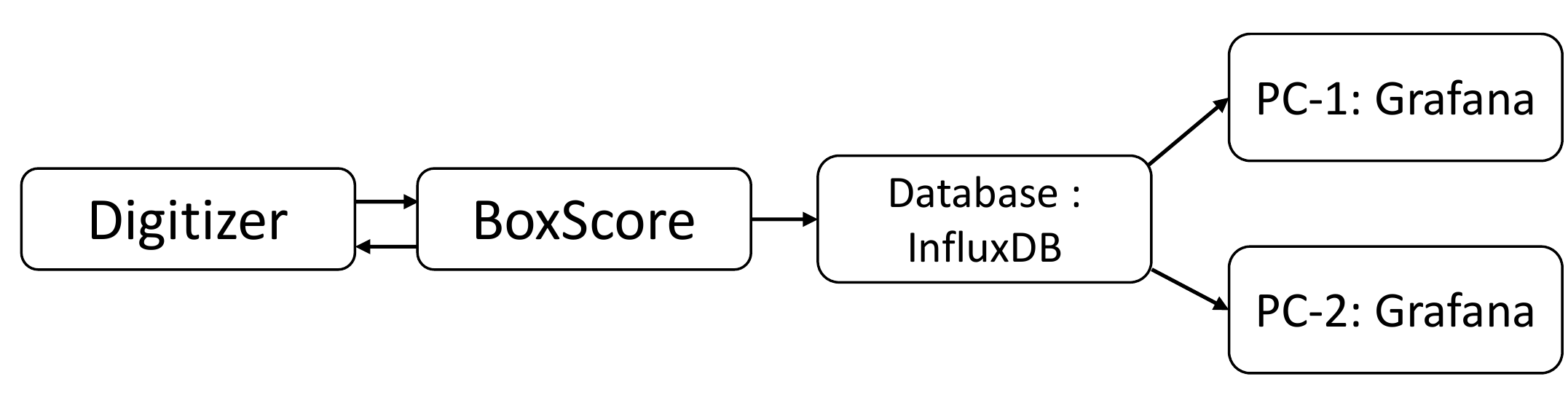}
\caption{\label{fig:Overview} An overview of the data acquisition system. \BoxScore~connects to digitizer for controling and reading data. \BoxScore~outputs rates of various isotopes into a database (local or via network). Other computers can acquire the rates from the database via network.}
\end{figure}
%----------------------------------------------

Fig.~\ref{fig:Overview} shows the overview of the data acquisition system. 
The digitizer is placed close to the detector system. 
A beam diagnosis station with \BoxScore~installed is placed near the digitizer. 
The database can be hosted locally at the beam diagnosis station or remotely in other computers. Grafana~\cite{grafana} can access and display the database on a dashboard. Any computers can view the dashboard via network connection. %\crhnote{why do other computers require Grafana to be installed to view? I thought it could be viewed anywhere if our network was setup correctly like at CERN.}

%The database can be displayed in multiple computers with Grafana installed.
%This allows for a centralized beam tuning station using data from beam diagnosis stations that are located remotely throughout the accelerator facility. 

%The program consists of several C++ Classes. The core is the \texttt{Digitizer} class that can access and readout the digitizer via USB or optical-link. The class stores the raw data (including waveform) and able to build events for a given time window. The \texttt{FileIO} class is used for saving the raw or built events into CERN Root file. The \texttt{GenericPlane} class is the generic canvas class for display histogram or count rate. Other canvas classes can be built upon the \texttt{GenericPlane} class for different applications. 

Here, we will first describe the architecture of \BoxScore, explaining the core components of the program. 
Next, its implementation at the ATLAS in-flight system is presented. Finally, the capability and performance of \BoxScore~is discussed.  

%######################################################
\section{Architecture of \BoxScore }

%\crhnote{It seems like you need a git link referenced somewhere with a fixed version of \BoxScore~that you used for this description as well as the performance tests. I am not sure if there are any argonne rules about publishing code developed here?? maybe it needs an official ANL github??}

The backbone of \BoxScore~consists of the \texttt{Digitizer}, the \texttt{GenericPlane}, and the \texttt{FileIO} classes. 
The role of each class is explained in the following.

The \texttt{Digitizer} class is for communication with a digitizer. 
It controls the digitizer by writing the registers in the digitizer, including parameters related to channel trigger threshold, input dynamic range, acquisition mode, the settings of the trapezoid filter, etc.
%\crhnote{need to be more explicit on items above, i.e., including parameters related to channel threshold settings, input signal dynamic range, etc.....}
The \texttt{Digitizer::ReadData()} method reads out data (pulse height and timestamp for each triggering signal) from the digitizer's buffer and saves the data into the computer's memory as raw data. 
In the ``mixed" mode, waveforms are also read and saved in addition to the pulse height and timestamp. 
The \texttt{Digitizer::BuildEvent()} method builds events from the raw data. 

The \texttt{GenericPlane} class is a canvas class that defines the canvas (using the \texttt{TCanvas} class in CERN ROOT) and plots histograms and graphs.
%The \texttt{GenericPlane::SetCanvasTitleDivision()} virtual method defines the division of the canvas.
The raw data, waveforms, or events can be analyzed and filled into histograms using the \texttt{GenericPlane::Fill()} virtual method. 
The \texttt{GenericPlane::Draw()} virtual method plots the histograms that depends on the division of the canvas.
The virtual methods are used so that other canvas classes can be inherited with minimum programming when other data processing algorithms or canvases are needed.
%For example, when other canvas or more complicated data processing is needed, a new class can be inherited from the \texttt{GenericPlane} class with minimum programming. 
%The \texttt{GenericPlane} class provides many key \texttt{virtual} functions for easy class inheritance, for example, Fill, XXXX, XXXXX. A TGraphicalCut ....

The \texttt{FileIO} class provides methods to save information into a CERN ROOT file. 
%Most CERN ROOT objects can be saved, such as \texttt{TMarco}, \texttt{Tree}, \texttt{TH1}, \texttt{TH2}, and \texttt{TGraph}. 
The digitizer settings are saved in \texttt{TMarco}. The events are saved in \texttt{TTree}. 
The raw data and waveforms are also saved in \texttt{TTree} when needed. 
The histograms (for example, PID plot) are saved in \texttt{TH1} or \texttt{TH2}. The graphs of the rates of selected isotopes are saved in \texttt{TGraph}. 

%----------------------FIGURE 2-----------------
\begin{figure}[ht]
\centering
%trim= left bottom right top
\includegraphics[trim=5.5cm 6.6cm 7.5cm 7.4cm, clip, width=9.cm]{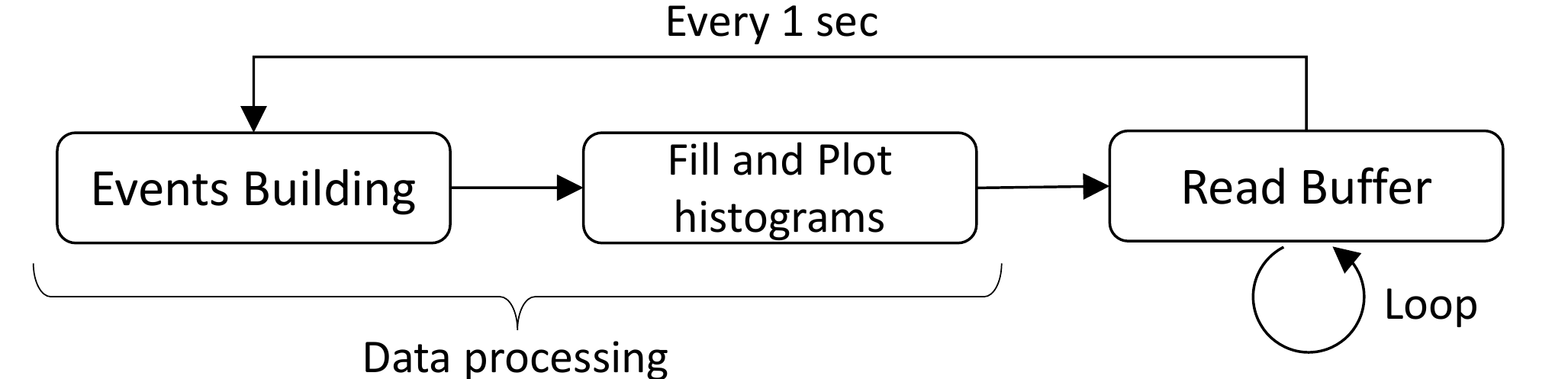}
\caption{\label{fig:flow} Flowchart of data acquisition in \BoxScore. Each second, events will be built from the collected raw data array and histograms will be plotted once the events were built. The raw data will be read from the digitizer buffer continuously.
%during the rest of a second.
}
\end{figure}
%----------------------------------------------

Fig.~\ref{fig:flow} shows the flowchart of \BoxScore~when the data acquisition is started. In order to realize real-time beam diagnosis, in each second, building events [\texttt{Digitizer::BuildEvent()}] is followed by filling [\texttt{GenericPlane::Fill()}] and plotting [\texttt{GenericPlane::Draw()}] histograms. 
%Processing data usually takes less than half second. 
The data processing (processes that do not read the digitizer's buffer) usually does not take a full second. 
After that, the data in the buffer (raw data) will be read and cleared from the digitizer, saved in the computer memory [\texttt{Digitizer::ReadData()}] repeatedly and constantly for the rest of the second to avoid the buffer being full and data lost. 
The memory for each channel is 5.12 MSample for both DT5730 and V17300 digitizers. If a signal consists of 4096 samples or 8192 ns, 1250 signals can be stored in a channel.
Each signal (or raw data) contains the pulse height (14 bits) and the timestamp (47 bits, 2 ns).
As the raw data is accumulated, it forms a 2-D array.
The pile-up signals, which have a timestamp but no pulse height, will be discarded.
%\crhnote{but are the counted? so you know how many you are losing?}
%\crhnote{what is the buffer size, how many events can it hold?}
%This setting guarantees the events will be built in each second. When the there is a lot raw data and the event-building takes long time, it reduces the time for data-reading, and reduces the times for next event-building. 

%Every second, events will be built based on the collected raw data and the time window (Fig.~\ref{fig:event_buidling}). 
%All raw data within the time window will be grouped and formed a single-event.

\begin{figure}[ht]
\centering
%trim= left bottom right top
\includegraphics[trim=3.4cm 4.5cm 3cm 7cm, clip, width=9.cm]{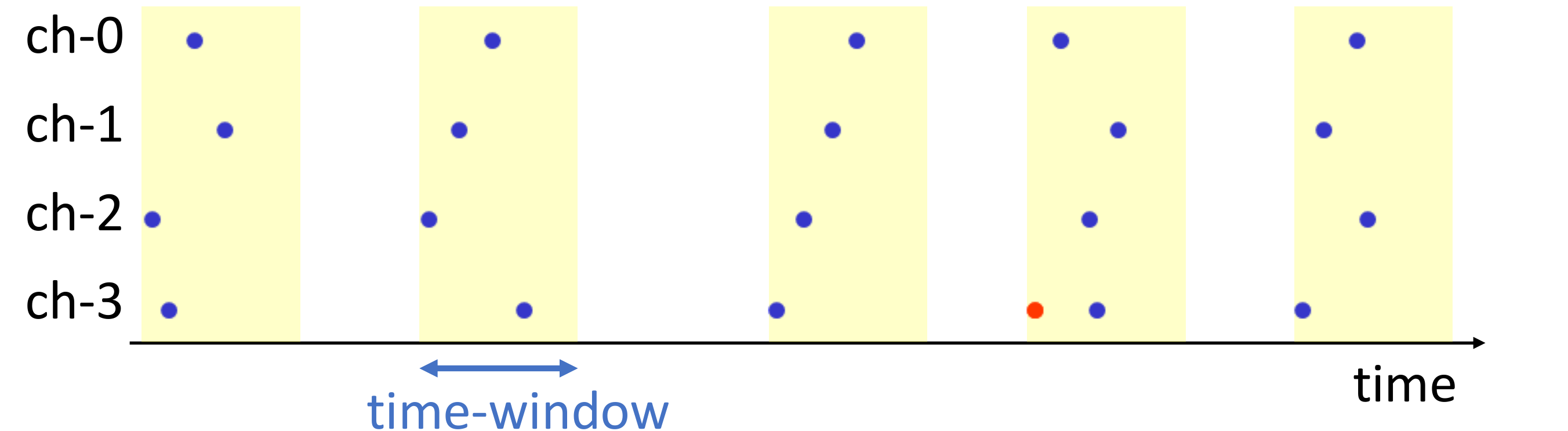}
\caption{\label{fig:event_buidling}Events building of \BoxScore. Only 4 channels are shown in here. The yellow boxes are the time window for event-building, it started with the earliest data after the last grouped data. If a channel was fired multiple times within a time window (multi-hit), only the last (blue) data will be stored and earlier (red) data will be dropped.
}
\end{figure}

In the event-building (Fig.~\ref{fig:event_buidling}), the raw data array is first sorted in ascending order according to the timestamp using the \texttt{TMath::BubbleLow()} method from CERN ROOT.
Next, the timestamp-sorted raw data will be grouped with the time window to form an event. 
For example, the sorted timestamp is denoted as $t_i$, where $i$ is the data index in the time-sorted raw data array. 
The first event starts from the $0$-th data with a timestamp $t_0$. When the $r$-th ($r>0$) data with a timestamp $t_r$ is the last data within the time window $T$, i.e. $t_r-t_0 < T$, data with timestamps $t_0, t_1,...,t_r$ will be grouped and form the first event. 
The second event starts from the $(r+1)$-th data with a timestamp $t_{r+1}$, when the $s$-th ($s>r+1$) data with a timestamp $t_s$ is the last data within the time window, i.e. $t_s-t_r < T$, then, data with timestamp $t_r, t_{r+1}, ..., t_s$ will be grouped as the second event. 
This process will continue until the last of the data. 
When the event-building has finished, the raw data array will be cleared except for the last group of the raw data. The last group of raw data will be left to the next raw data array, to avoid incomplete data collecting from the digitizer during the data processing stage. 
If there are multiple raw data from a single channel (which is called multi-hit) within the time window, only the last raw data from that channel will be saved.%\crhnote{again, is it counted though to keep track some how?}

%\crhnote{the statement above and Fig. 3 talk about pile-up. The figure shows far more pile-up then we should see or have. If you are going to talk about this, then \BoxScore needs to have reporting of the amount of pile-up. But in reality we have very little with only 1-2kHz rates. I think the figure is misleading unless you have real data to prove otherwise.}
%----------------------FIGURE 3-----------------
%----------------------------------------------

The time required for event-building is proportional with the number of data in the raw data array. %\crhnote{does size = number of raw data events. I would be careful with using the word size because to me it refers to an amount of memory or storage, i.e. bytes}
In order to avoid the event-building taking longer than a second, a size limit of 100 thousand was set for the raw data array.
This corresponds to a sum of 100 kHz triggers from all channels theoretically. %\crhnote{is this near to any limits in the internal digitizer buffer size? If so, events could fall off the buffer on the digitizer.}

The waveform is also stored in the \texttt{Digitizer} class when the data acquisition mode is set to ``mixed". 
When the \texttt{Digitizer::ReadData()} is called, waveforms from all channels are being read in this mode. 
In each channel, many waveforms could be taken from the digitizer's buffer.
However, only the earliest waveform is saved and stored, later waveforms are discarded.
When using \BoxScore~as an oscilloscope, the waveforms are displayed after \texttt{Digitizer::ReadData()} is called, and the \texttt{Digitizer::BuildEvent()} will not be called.

%######################################################
\section{Implementation in ATLAS in-flight system and RAISOR}

\BoxScore~was implemented as an essential part of the ATLAS in-flight system for isotope identification and beam transport optimization (beam tuning) in real-time. 
There are many beam diagnostic planes (or locations, stations) along the beamline from the ATLAS in-flight system to the target destination, HELIOS~\cite{Lighthall10}, the beam line used in this study. For example, in this case there were diagnostic planes at the RAISOR focal plane, at the entrance of HELIOS, at the HELIOS target position, and at the exit of HELIOS. 
At each plane, various silicon detectors were used and various canvas classes were derived from the \texttt{GenericPlane} class to display the required histograms and plots. 

\begin{figure}[ht]
\centering
%trim= left bottom right top
\includegraphics[trim=0cm 0cm 0cm 0.7cm, clip, width=8.5cm]{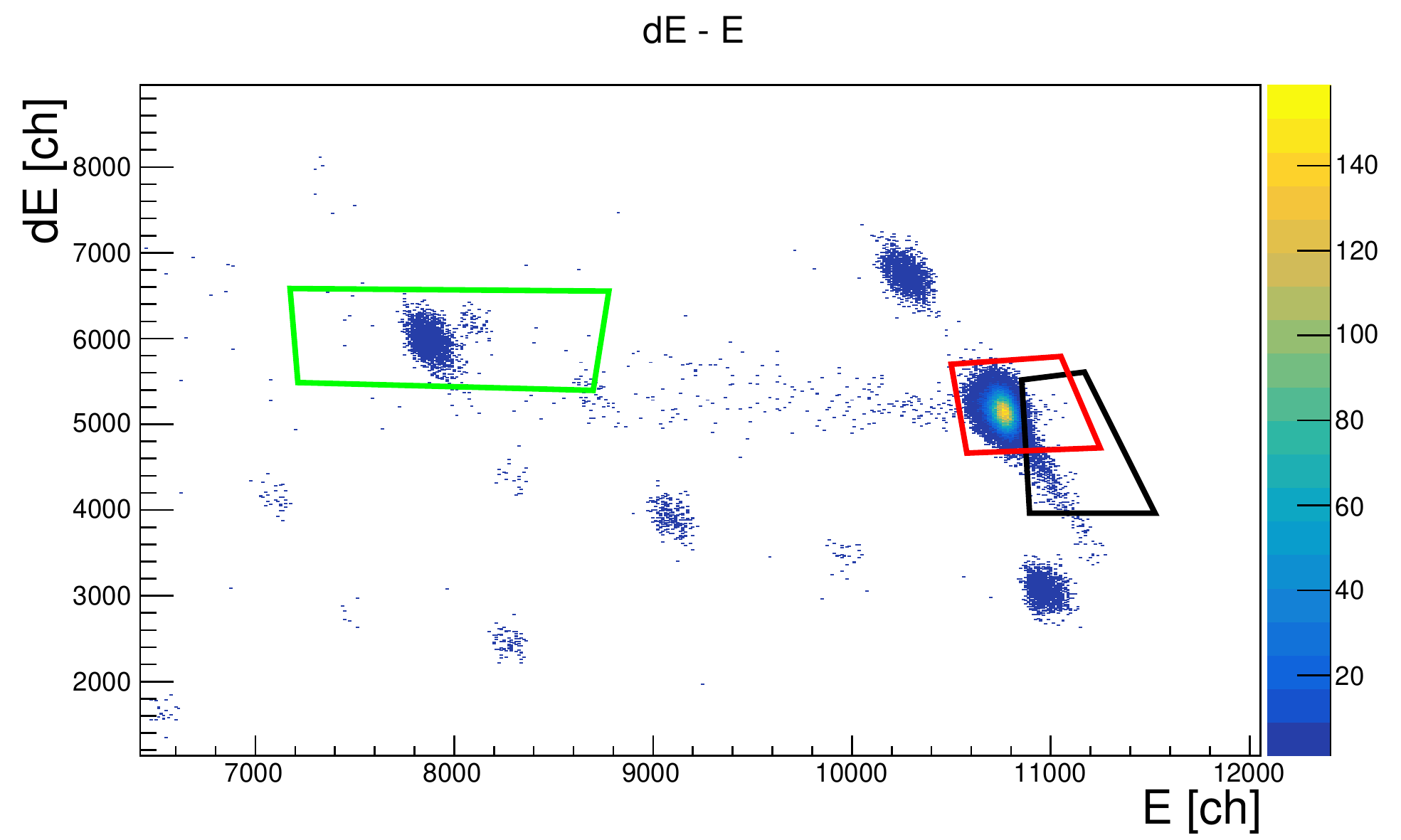}
\caption{\label{fig:dE-E_exist} A screenshot of a PID plot from \BoxScore. The beam was produced by the \iso{15}{N}($d$,$p$)\iso{16}{N} reaction. \iso{16}{N} is inside the red box and \iso{15}{N} is inside the green box. The black box is the tail of \iso{16}{N}.}
\end{figure}

\begin{figure}[ht]
\centering
%trim= left bottom right top
\includegraphics[trim=0cm 0cm 0cm 0.6cm, clip, width=8.5cm]{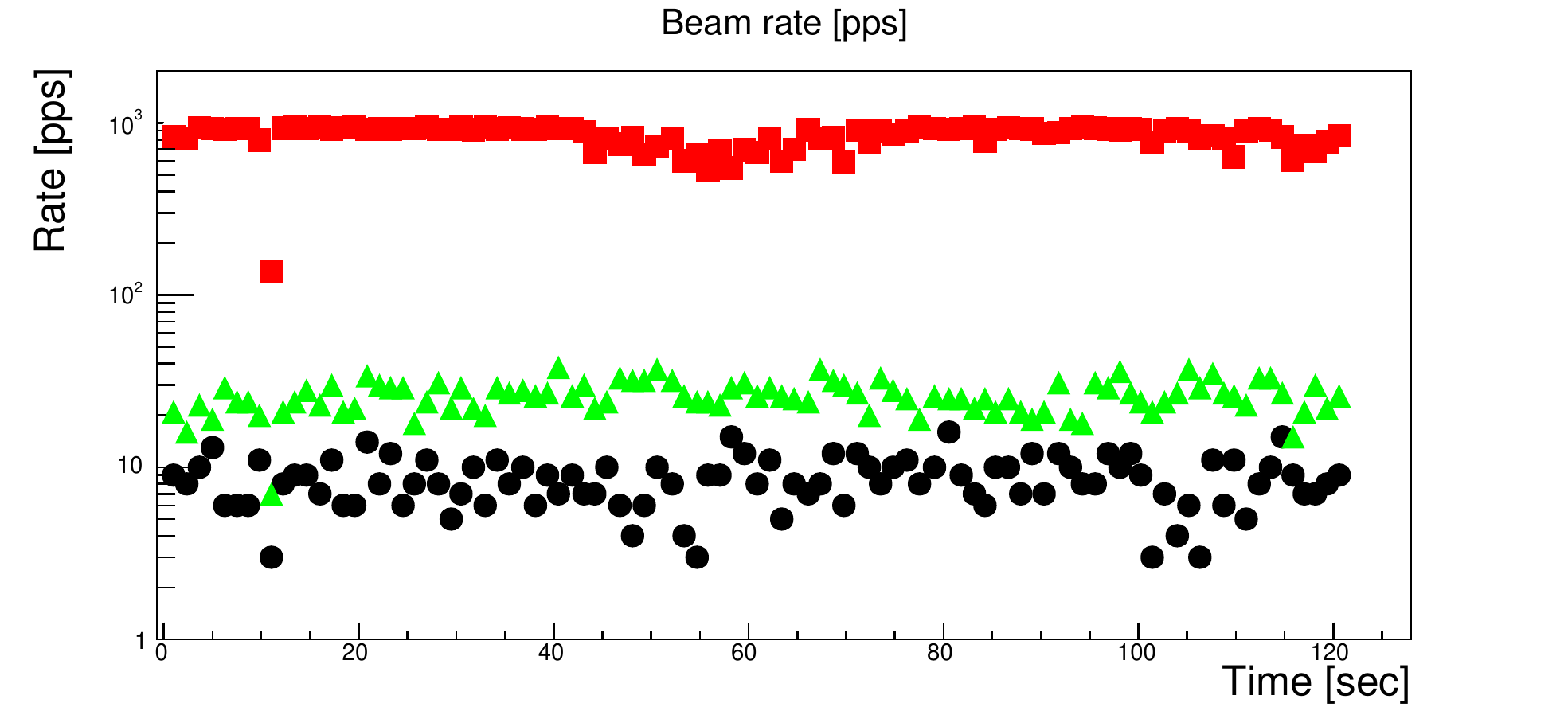}
\caption{\label{fig:rate} The isotope rates from the boxes (selections) in Fig.~\ref{fig:dE-E_exist}.}
\end{figure}

A typical histogram is a $\Delta E-E$ plot used for PID. Fig.~\ref{fig:dE-E_exist} shows the PID of a \iso{16}{N} beam produced by the ATLAS in-flight system using the \iso{15}{N}($d$,$p$)\iso{16}{N} reaction in inverse kinematics.
Graphical selection can be created online (the colored polygons in Fig.~\ref{fig:dE-E_exist}). Once selections are created, the rates of the selections will be shown in a rate graph (Fig.~\ref{fig:rate}), and the selections will be saved in the output CERN ROOT file. 

%----------------------FIGURE 4-----------------
%----------------------------------------------

%----------------------FIGURE 5-----------------
%----------------------------------------------

\begin{figure}[ht]
\centering
%trim= left bottom right top
\includegraphics[trim=0cm 0.35cm 0cm 0cm, clip, width=6.cm]{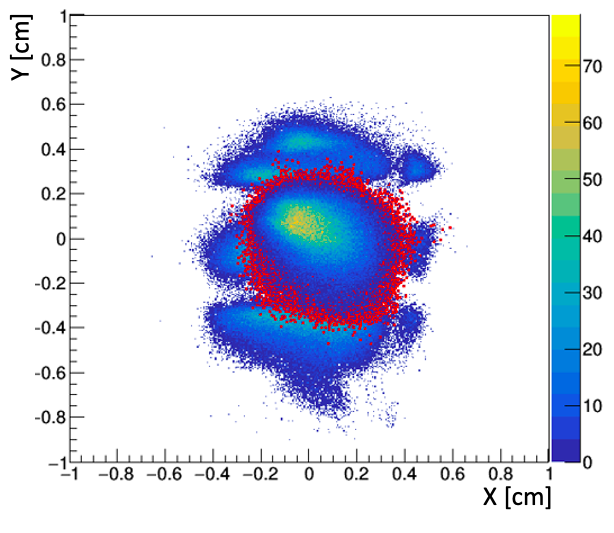}
\caption{\label{fig:beam_profile} A perpendicular beam profile using a position sensitive silicon detector at the HELIOS target position. The red shaded region at the center of the plot highlighted an isotope that is selected in a PID plot. The data is taken form other experiment, which is different from Fig.~\ref{fig:dE-E_exist}.}
\end{figure}

A double-sided position sensitive silicon detector~\cite{MSPSD} could also be installed to measure the beam profile and PID at the same time. 
Fig.~\ref{fig:beam_profile} shows an example of a beam profile that is perpendicular (x-y plane) to the beam direction (z-axis). 
It shows many beam spots corresponding to various isotopes. 
The red shaded region highlighted an isotope that was selected in a PID plot (not shown). 
The position sensitive detector had five outputs, so that building an event from 5-channels and calculation of the position are required. 
Building events from other channels and others complicated calculations can be achieved by creating a new canvas class inherited from the \texttt{GenericPlane} class. 
For example, a 4-channel event-building and trace-analysis algorithm was tested to construct a beam profile from a microchannel plate detector~\cite{MCP}. %\crhnote{more info needed in particular for the position sensitive detector, i.e. what positions transverse? resolution, usefulness? There has not been mention of a need for position before this.}

%----------------------FIGURE 6-----------------
%----------------------------------------------

%######################################################
\section{Performance}
%\crhnote{Again, would be good to link to a specific version of \BoxScore that the tests were done with...}

\BoxScore~v2.0, released on Oct 29\textsuperscript{th}, 2021, has been used in the testing described below~\cite{BoxScore}. 
Execution of \BoxScore~requires less than \mbox{200 MB} of RAM. 
The performance of \BoxScore~was tested using a \mbox{Ubuntu 18} machine with Intel\textsuperscript{\textregistered}~Core\texttrademark~ i5-8250U CPU and 4 GB RAM, connected to a DT5730 digitizer via USB cable. 
Input signals were generated by a CAEN DT5810 Fast Digital Detector Emulator. 
The signal has a rise time of 20 ns and a decay time of 50 $\mu$s with a fixed amplitude of 1 V that is sufficient to trigger the digitizer. 
%The theoretical maximum rate for non-pile up signal is $\approx$10 kHz. \crhnote{10 kHz where? based on what?}
The sample size for the digitizer was set to be 4096 Samples or 8912 ns, so that the buffer can store 1250 signals for each channels.
The coincident time window was 400 ns.
The trapezoid settings for the digitizer are 100 ns for the rise-time, 200 ns for the flat-top, 50 $\mu$s for the pole-zero, 150 ns for the peaking time, and 100 ns for the peak-hold off. 
Under this trapezoid setting, the theoretical maximum pile-up free trigger rate is 2 MHz. 
In addition, the database output was disabled (Fig.~\ref{fig:Overview}) and also no network connection to a remote database host for a reason that will be explained later. 
The maximum rate (that the rates from the signal emulator is the same to the measured trigger rate from the digitizer) in the ``list" and ``mixed" modes for single channel, 2 channels, and 4 channels are listed in Table~\ref{table:rate}.

%----------------------------------------------------------------------------
\begin{table}[ht]
\centering
\caption{\label{table:rate} Performance for the DT5730 digitizer. The sample size was 4096 samples.}
\newcommand\T{\rule{0pt}{3ex}}
\newcommand \B{\rule[-1.8ex]{0pt}{0pt}}
%\begin{ruledtabular}
\begin{tabular}{c  c  c  c  c}
\hline
\hline
 & 1-channel & 2-channel & 4-channel  \\
\hline
%\\[-0.7em]
Event-building time & $\sim30$ ms & $\sim60$ ms & $\sim140$ ms \\
Max. rate (``list")  & 5.5 kHz & 4.7 kHz & 3.8 kHz\\
Max. rate (``mixed") & 15 Hz & 15 Hz & 15Hz\\
\hline
\end{tabular}
%\end{ruledtabular}
\end{table}
%----------------------------------------------------------------------------

The time for the data processing mainly depends on the rate of input signal and number of input channels (Fig.~\ref{fig:flow}). 
%\crhnote{As i understand it, essentially everything needs to happen under 1 sec or before the next data read right? These dont' add to 1 sec so why the limits? what else is involved?}
The time duration for the data saving and plotting histograms is about $\sim$70 ms and $\sim$90 ms, respectively, disregarding the number of channels. 
But the event-building time increases from $\sim$30 ms for 1 channel at 5.5 kHz, to 60 ms for 2 channels at 4.7 kHz, to $\sim$140 ms for 4 channels at \mbox{3.8 kHz}. 
The database output to a remote host via network depends on the network environment and would be very time consuming. %\crhnote{i am not sure what this means? do you mean writign to a remote database vs. local database is slow?}
If any of the above processes take a lot of time so that the buffer could have been filled before the Digitizer::ReadData() method is called, the maximum rate will reduce. %\crhnote{ this needs to be quantified, a lot of time doesn't mean anything, i.e. $>100$~ms or $>$500~ms...}, %that will reduce the amount of time for the \texttt{Digitizer::ReadData()} method being called. \marker{Since there is bandwidth limit for USB cable, the amount of buffer can be read is limited,} and eventually 
For example, if the total time for data processing is 250 ms, the buffer of 1250 signals will be full for a 5 kHz signal.
Since the number of signals that can be stored in the buffer can be adjusted by the sample size, the maximum rate could be increased by reducing the sample size.
The maximum rate for waveform (or in ``mixed") mode is limited by the algorithm that only the earliest waveform will be stored in the \texttt{Digitizer::ReadData()} method (Fig.~\ref{fig:flow}). %\crhnote{again I am not sure what this means, you are 'choosing' only to read the first one right? so its not really limited to this rate, it is fixed to this rate.}

%######################################################
\section{Summary and outlook}

\BoxScore~was developed to meet the demands of real-time beam diagnosis and monitoring. 
The program is an essential component for developing in-flight beams using the ATLAS in-flight system now.
It builds events from multiple channels every second, provides the PID, and is able to output the count rates for selected isotopes. 
\BoxScore~contains 3 key classes: \texttt{Digitizer}, \texttt{GenericPlane}, and \texttt{FileIO}.
The program was tested with a single CAEN DT5730 digitizer via USB connection and it should be compatible with all x730 series digitizers with DPP-PHA firmware. 
It is a lightweight, portable program that could be expanded to handle many nuclear physics applications that required real-time event-building. %\crhnote{I am not sure about this, it is not ready to read all kinds of experiments, it has been developed for a few channels at certain rates. I would say it could be expanded to handle a number of nuc. phys. applications.}
There are many things that could be developed in \BoxScore, for example, supporting the Pulse Shape Discrimination firmware, supporting and synchronizing multiple digitizers, increasing the maximum rate by algorithm optimization, and integrate into NSCLDAQ~\cite{NSCLDAQ}, etc. 
\BoxScore~is open-source and constantly evolving \cite{BoxScore}. %Any feedback and suggested modifications are welcome. \crhnote{I would remove this and replace with the github link that implies all of these words.}

%######################################################
\section{Acknowledgement}
The author would like to acknowledge the support and operations staff at ATLAS and fruitful discussions with C.~R.~Hoffman and G.~L.~Wilson. This research used resources of Argonne National Laboratory’s ATLAS facility, which is a Department of Energy Office of Science User Facility. This material is based upon work supported by the U.S. Department of Energy, Office of Science, Office of Nuclear Physics, under Contract No. DE-AC02-06CH11357. 

%%\appendix

%\section{Sample Appendix Section}
%Lorem ipsum dolor sit amet, consectetur adipiscing elit, sed do eiusmod tempor section \ref{sec:sample1} incididunt ut labore et dolore magna aliqua. Ut enim ad minim veniam, quis nostrud exercitation ullamco laboris nisi ut aliquip ex ea commodo consequat. Duis aute irure dolor in reprehenderit in voluptate velit esse cillum dolore eu fugiat nulla pariatur. Excepteur sint occaecat cupidatat non proident, sunt in culpa qui officia deserunt mollit anim id est laborum.

%\bibliographystyle{elsarticle-num} 
%\bibliography{cas-refs}


\end{document}